\documentclass[12pt]{article}
\usepackage{graphicx}

\newcommand\pubnumber{}
\newcommand\pubdate{\today}

\def\bari{$^1$Dipartimento Interateneo di Fisica ``Michelangelo Merlin,''\\
Via Amendola 173, 70126 Bari, Italy \\ 
$^2$Istituto Nazionale di Fisica Nucleare (INFN), Sezione INFN di Bari,\\
Via Orabona 4, 70126 Bari, Italy}

\def\Title#1{\begin{center} {\Large #1 } \end{center}}
\def\Author#1{\begin{center}{ \sc #1} \end{center}}
\def\Address#1{\begin{center}{ \it #1} \end{center}}

\newcommand\pubblock{\rightline{\begin{tabular}{l} \pubnumber\\
         \pubdate  \end{tabular}}}
\newenvironment{Abstract}{\begin{quotation}  }{\end{quotation}}
\newenvironment{Presented}{\begin{quotation} \begin{center} 
             PRESENTED AT\end{center}\bigskip 
      \begin{center}\begin{large}}{\end{large}\end{center} \end{quotation}}
\def\Acknowledgements{\bigskip  \bigskip \begin{center} \begin{large}
             \bf ACKNOWLEDGEMENTS \end{large}\end{center}}

\def\beq{\begin{equation}}
\def\eeq#1{\label{#1}\end{equation}}
\def\eeqn{\end{equation}}

\def\beqa{\begin{eqnarray}}
\def\eeqa#1{\label{#1}\end{eqnarray}}
\def\eeqan{\end{eqnarray}}

\let\bar=\overbar

\def\Dslash{\not{\hbox{\kern-4pt $D$}}}
\def\dslash{\not{\hbox{\kern-2pt $\del$}}}

\def\msb{{\bar{\ssstyle M \kern -1pt S}}}

\begin{document}
\begin{titlepage}
\pubblock

\vfill
\Title{Short- and long-baseline sterile neutrino phenomenology}
\vfill
\Author{Antonio Palazzo$^{1, 2}$}
\Address{\bari}
\vfill
\begin{Abstract}
Several anomalies observed in short-baseline neutrino experiments indicate that 
the standard 3-flavor framework may be incomplete and point towards the existence
of light sterile neutrinos. Here, we present a concise review of the status of the neutrino
oscillations within the 3+1 scheme, which is a minimal extension of the standard 3-flavor framework
with one sterile neutrino species.  We emphasize the potential role of LBL experiments in the searches 
of CP violation connected to sterile neutrinos and their complementarity with the SBL experiments.
\end{Abstract}
\vfill
\begin{Presented}
NuPhys2016, Prospects in Neutrino Physics\\
Barbican Centre, London, UK, December 12--14, 2016
\end{Presented}
\vfill
\end{titlepage}
\def\thefootnote{\fnsymbol{footnote}}
\setcounter{footnote}{0}

\section{Introduction}

A long series of experiments performed in the last twenty years has established that neutrinos
are massive and mix. Although the 3-flavor framework has been nailed out as the only one able
to describe simultaneously all data collected in solar, atmospheric, reactor and accelerator neutrino
experiments, some ``anomalous'' results have been reported in short-baseline (SBL) neutrino oscillation measurements, 
which cannot be explained in such a scheme (see~\cite{Palazzo:2013me} for a recent review).
The most popular interpretation of these anomalies involves new light sterile neutrinos 
(with mass in the eV range) which mix with the three ordinary ``active" species.
  
In the simplest scenario, dubbed as 3+1 scheme, only one sterile neutrino species is introduced.
The fourth mass eigenstate $\nu_4$ is separated from the standard ``triplet'' $(\nu_1, \nu_2,\nu_3)$
by a large ($\sim$\,1\,eV$^2$) squared-mass, giving rise to the hierarchical  spectrum
$|\Delta m^2_{12}| \ll  |\Delta m^2_{13}| \ll |\Delta m^2_{14}| $. The $4\times4$ mixing matrix can be parametrized as~\cite{Klop:2014ima} 
\begin{equation}
\label{eq:U}
U =   \tilde R_{34}  R_{24} \tilde R_{14} R_{23} \tilde R_{13} R_{12}\,, 
\end{equation} 
where $R_{ij}$ ($\tilde R_{ij}$) is a real (complex) $4\times4$ rotation in the ($i,j$) plane. 
The complex rotations depend on the CP-phases $\delta_{ij}$.

By construction,  the $3+1$ scheme predicts sizable oscillation effects at the short baselines
where the new frequency $\Delta_{14} = \Delta m^2_{14} L/4E$ ($L$ being the baseline and $E$
the neutrino energy) is of order one.  
However, sterile neutrinos may leave their signs also in non-short-baseline
experiments. In these setups the new oscillations get averaged because they are very fast
($\Delta_{14} \gg 1 $) and the manifestation of active-sterile oscillations is more subtle. 

In the solar sector, for example, the admixture of the electron neutrino with the $\nu_4$ mass eigenstate 
(parametrized by the matrix element $U_{e4}$) leads to small deviations 
from the unitarity of the $(\nu_1, \nu_2,\nu_3)$ sub-system (see~\cite{Palazzo:2011rj,Palazzo:2012yf,Giunti:2009xz}).
In the atmospheric neutrinos, as first evidenced in~\cite{Nunokawa:2003ep}, at very high [$O$(TeV)] energies,
a striking MSW resonant behavior is expected, which leads to a distortion of the zenith angle distributions. 
Finally, as first evidenced in~\cite{Klop:2014ima}, sterile neutrino oscillations can be probed also 
in the long-baseline (LBL) accelerator experiments, where they give rise to new interference phenomena.
This circumstance is of particular interest in view of the world-wide program of new LBL facilities. 

In what follows we concisely describe the  anomalous SBL results and describe the potential role of LBL
experiments in the searches of CP violation (CPV) related to sterile neutrinos.

\section{The LSND and MiniBoone anomalies}

Accelerator experiments with baselines of a few tens of meters and neutrino energies
of a few tens of MeV can effectively probe neutrino oscillations occurring at $\Delta m^2 \sim 1\,\mathrm{eV}^2$.  
Their results are usually interpreted introducing a new mixing angle $\theta$ and 
a new mass-squared difference $\Delta m^2$. In the 3+1 framework the following identifications are valid:
 $\Delta m^2 \equiv \Delta m^2_{14}$ and $\sin^2 2\theta \equiv 4  |U_{e4}|^2 |U_{\mu4}|^2$.
The anomalous result recorded at the LSND experiment~\cite{Athanassopoulos:1996jb}
was the first piece of data pointing towards light sterile neutrinos. Such an experiment, 
designed to study $\bar\nu_\mu\to \bar\nu_e$ transitions, detected an excess 
of electron antineutrino events a the $\sim3.8\sigma$ level.
The experiment  KARMEN~\cite{Armbruster:2002mp}, having a design very similar to LSND, 
observed no such a signal, but could not rule out entirely the mass-mixing parameter region allowed by LSND.
The experiment MiniBooNE~\cite{AguilarArevalo:2012va},  sensitive both to $\nu_\mu\to \nu_e$  and  $\bar\nu_\mu\to \bar\nu_e$
transitions, seems to lend support to the LSND finding.

An independent test of the LSND and MiniBooNE anomalies has been recently carried out at the long-baseline
experiments ICARUS~\cite{Antonello:2012pq} and OPERA~\cite{Agafonova:2013xsk}.
In these experiments, due to the high energy of the beam ($<E>$\,\,$ \sim 17$~GeV), the 3-flavor effects induced by 
non-zero $\theta_{13}$ are negligible. As a result, the experiments are sensitive to sterile neutrino
oscillations, although these are completely averaged out due to the high value $L/E \sim  36.5$~m/MeV,
and appear as an enhancement of the expected rate of events.
Both collaborations have performed the analysis in an effective 2-flavor description.
However, in~\cite{Palazzo:2015wea} it has been pointed out that in the 3+1 scheme important corrections arise due
to the presence of a new genuinely 4-flavor interference term in the transition probability.
In addition, in the 4-flavor framework the  $\nu_e$ beam contamination is not a fixed
quantity like in the 2-flavor scheme. A consistent 4-flavor analysis~\cite{Palazzo:2015wea} of ICARUS and OPERA
leads to a substantial weakening (by a factor of $\sim$ 3) of the upper bounds on the sterile
neutrino mixing. ICARUS and OPERA are not sensitive enough to rule out the
mass-mixing region preferred by LSND and MiniBooNE, and can only restrict 
the allowed region to values of $\sin^2 2\theta <  few \times 10^{-2}$.

\section{The reactor and gallium anomalies}

The re-calculations of the reactor antineutrino spectra performed 
in~\cite{Mueller:2011nm,Huber:2011wv} have given new momentum to the study of light sterile neutrinos.
These calculations indicate fluxes which are $\sim 3.5\%$ higher than previous estimates and have raised 
the so-called reactor antineutrino anomaly~\cite{Mention:2011rk}. In fact, adopting these fluxes, the SBL 
reactor measurements show a deficit with respect to the theoretical expectations.

The recent high statistics measurements of the antineutrino spectra performed by 
Daya Bay~\cite{An:2016srz}, Double Chooz~\cite{Schoppmann:2016iww} and RENO~\cite{Seo:2016uom}
(and more recently also by NEOS~\cite{Ko:2016owz}) have evidenced an unexpected bump around $5$ MeV
in the prompt energy spectrum, deviating from the predictions at the $\sim4 \sigma$ level.
The bump structure appears to be similar at the near and far detectors and is positively
correlated with the reactor power. This strongly disfavors a possible explanation in terms of 
new-physics (for example super-light sterile neutrinos~\cite{Palazzo:2013bsa}). 
This unexpected feature of the spectrum evidences that our understanding of the reactor spectra is incomplete
and reinforces the case for a revision of the current reactor flux predictions.
To this regard, it has been recently pointed out~\cite{Giunti:2016elf} (see also~\cite{An:2017osx}),
that if one hypothesizes  that the reactor anomaly is not due to active-sterile neutrino oscillations, it
can be explained entirely by a miscalculation of the $^{235}{\mathrm U}$ reactor antineutrino flux. 

An apparently unrelated deficit has been found in the solar neutrino experiments
GALLEX and  SAGE~\cite{Abdurashitov:2005tb} using high intensity radioactive sources. 
The statistical significance of the deficit fluctuates around the $3\sigma$ level slightly 
depending on the assumptions made on the theoretical estimate of the cross section
 $\nu_e + ^{71}\!{\mathrm {Ga}} \to ^{71}\!{\mathrm {Ge}} + e^-$. While the anomaly may represent a signal of new physics, 
both a systematic error in the ${\mathrm {Ge}}$ extraction efficiency or in the theoretical estimate of 
the cross-section remain possible alternative explanations.

Both the reactor and gallium anomalies can be interpreted in terms of a phenomenon of
electron neutrino disappearance driven by sterile neutrino oscillations. 
In an effective 2-flavor scheme the results can be described by a new mass-squared difference 
$\Delta m^2$  and an effective mixing angle $\theta$.
In a 3+1 framework the following 
 identifications hold: $\Delta m^2 \equiv \Delta m^2_{14}$ and 
 $\sin^2 2\theta \equiv 4 |U_{e4}|^2 (1- |U_{e4}|^2)$. 
 The simultaneous explanation of both anomalies requires values of $\Delta m^2 \sim 1.7 \, \mathrm{eV}^2$ 
 and $\sin^2 2\theta \sim 0.1$ (see~\cite{Gariazzo:2017fdh}). The inclusion of the recent results from 
the reactor experiment NEOS~\cite{Ko:2016owz} tends to shift downward the best fit value of the mixing angle to 
$\sin^2 2\theta \sim 0.08$ (see~\cite{Gariazzo:2017fdh}). It should be noted that, if the best fit point
lies in the region indicated by NEOS, the detection of a signal at the future SBL experiments will be more
challenging than originally envisaged. 

The results of NEOS deserve some further comment. A raster scan analysis made by
the collaboration excludes the hypothesis of oscillations at the 90\% C.L.  On the other hand, 
the expansion of the $\Delta \chi^2$ evidences a preference for sterile neutrino oscillations at roughly
$2\sigma$ in 2 d.o.f. (standard 3-flavor case disfavored at $\Delta \chi^2 = 6.5$). In addition, the region 
of parameters identified by NEOS lies inside the region allowed by all the current SBL data (see~\cite{Gariazzo:2017fdh}).
We think that  these two findings are intriguing and deserve more attention.

\section{Sterile neutrinos at long-baseline experiments}

The short-baseline experiments are without doubt the best place where to look for sterile neutrinos
and certainly, if a breakthrough will come, it will take place at a SBL experiment. However,
the SBL experiments have an intrinsic limitation which would impede to further study
the properties of the 3+1 scheme. In particular, they are insensitive to the 
three CP-violation phases involved in such a scheme. In fact, CP-violation is a genuine 3-flavor
phenomenon, whose observation requires the sensitivity to the {\em interference} between at least
{\em two independent} oscillation frequencies. In a SBL experiment only the new largest oscillation frequency
($\Delta_{14} \sim 1$) is visible, while both the atmospheric and the solar splittings are substantially 
unobservable ($\Delta_{13} \simeq \Delta_{12} \simeq 0$).
Therefore, this class of experiments is blind to CP-violation effects.%
 \footnote{In the $3+N_{s}$ schemes with $N_{s} >1$,  CPV could be observed at SBL experiments. 
 However, these setups can probe only a limited number of all the CP phases involved in the model.
 In contrast,  LBL experiments have access to all such phases. For example, in the 3+2 scheme, 
 the SBL experiments are sensitive only to one CP-phase over a total of five observable CP-phases. }
Other kinds of experiments are necessary to access the CP violation induced by sterile neutrinos.
We are lucky because these experiments already exist. 
We are talking of the LBL experiments, both those already operational and the planned ones.
 As a matter of fact, although such experiments were 
originally designed to seek the standard CP-phase $\delta$, they are also capable to provide information
about other sources of CP violation. This is not obvious a priori and it is true only because, as we will see below,
a new interference term arises in the presence of sterile neutrinos, which has exactly the same order
of magnitude of the standard 3-flavor interference term. 

We recall that the LBL experiments, when working in the $\nu_\mu \to \nu_e$ (and $\bar \nu_\mu \to \bar \nu_e$) 
appearance channel are sensitive to the 3-flavor CPV because, at long baselines, the $\nu_\mu \to \nu_e$ 
transition amplitude develops an interference term between the atmospheric ($\Delta m^2_{13}$-driven)
and the solar ($\Delta m^2_{12}$-driven) oscillations, which depends on the CP-phase $\delta$.
As first evidenced in~\cite{Klop:2014ima}, in the presence of sterile neutrinos a new interference term arises, 
which depends  not only from $\delta \equiv \delta_{13}$ but also from one new CP-phase ($\delta_{14}$). 
From the discussion made in~\cite{Klop:2014ima} (see also~\cite{Palazzo:2015gja}), it emerges that the 
transition probability can be approximated as the sum of three terms
\begin{eqnarray}
\label{eq:Pme_4nu_3_terms}
P^{4\nu}_{\mu e}  \simeq  P^{\rm{ATM}} + P^{\rm {INT}}_{\rm I}+   P^{\rm {INT}}_{\rm II}\,.
\end{eqnarray}
The first term represents the positive-definite atmospheric transition probability, 
while the other two terms (which can assume both positive and negative values) are
due to interference. The first of them is related to the well-known standard atmospheric-solar
interference, while the second is driven by the atmospheric-sterile interference.
We now observe that the transition probability depends on the three small mixing angles $\theta_{13}, \theta_{14}, \theta_{24}$,
and that it occurs that these mixing angles have the same order of magnitude $\epsilon \sim 0.15$.
In addition, we note that the ratio of the solar and atmospheric squared-mass splittings  
$\alpha \equiv \Delta m^2_{12}/ \Delta m^2_{13} \simeq \pm 0.03$, which
is of order $\epsilon^2$. Keeping terms up to the third order, in vacuum, one finds~\cite{Klop:2014ima} 
\begin{eqnarray}
\label{eq:Pme_atm}
 &\!\! \!\! \!\! \!\! \!\! \!\! \!\!  P^{\rm {ATM}} &\!\! \simeq\,  4 s_{23}^2 s^2_{13}  \sin^2{\Delta}\,,\\
 \label{eq:Pme_int_1}
 &\!\! \!\! \!\! \!\! \!\! \!\! \!\! \!\! P^{\rm {INT}}_{\rm I} &\!\!  \simeq\,   8 s_{13} s_{12} c_{12} s_{23} c_{23} (\alpha \Delta)\sin \Delta \cos({\Delta + \delta_{13}})\,,\\
 \label{eq:Pme_int_2}
 &\!\! \!\! \!\! \!\! \!\! \!\! \!\! \!\! P^{\rm {INT}}_{\rm II} &\!\!  \simeq\,   4 s_{14} s_{24} s_{13} s_{23} \sin\Delta \sin (\Delta + \delta_{13} - \delta_{14})\,,
\end{eqnarray}
where $s_{ij} \equiv \sin \theta_{ij}$, $c_{ij} \equiv \cos \theta_{ij}$ and  $\Delta \equiv  \Delta m^2_{13}L/4E$ is the 
atmospheric oscillating frequency. 
The matter effects slightly modify the transition probability leaving the picture described above almost 
unaltered (see~\cite{Klop:2014ima,Palazzo:2015gja} for details).

Remarkably, for typical values of the mixing angles preferred by the current global 3+1 
fits~\cite{Gariazzo:2017fdh}, the amplitude of the new interference
term is almost identical to that of the standard one. 
As a consequence, one expects some sensitivity of the LBL experiments NO$\nu$A~\cite{Adamson:2017gxd}
and T2K~\cite{Abe:2017uxa} to the non-standard CP-phase $\delta_{14}$. Therefore, these two experiments and
their constraints on the CP-phase $\delta_{14}$, should be included in any accurate analysis
of the 3+1 scheme. This has been done in the work~\cite{Capozzi:2016vac}, where
a joint analysis of SBL and LBL data has been presented for the first time.

\begin{figure}[t!]
\vspace*{0.0cm}
\hspace*{3.1cm}
\includegraphics[width=9.5 cm]{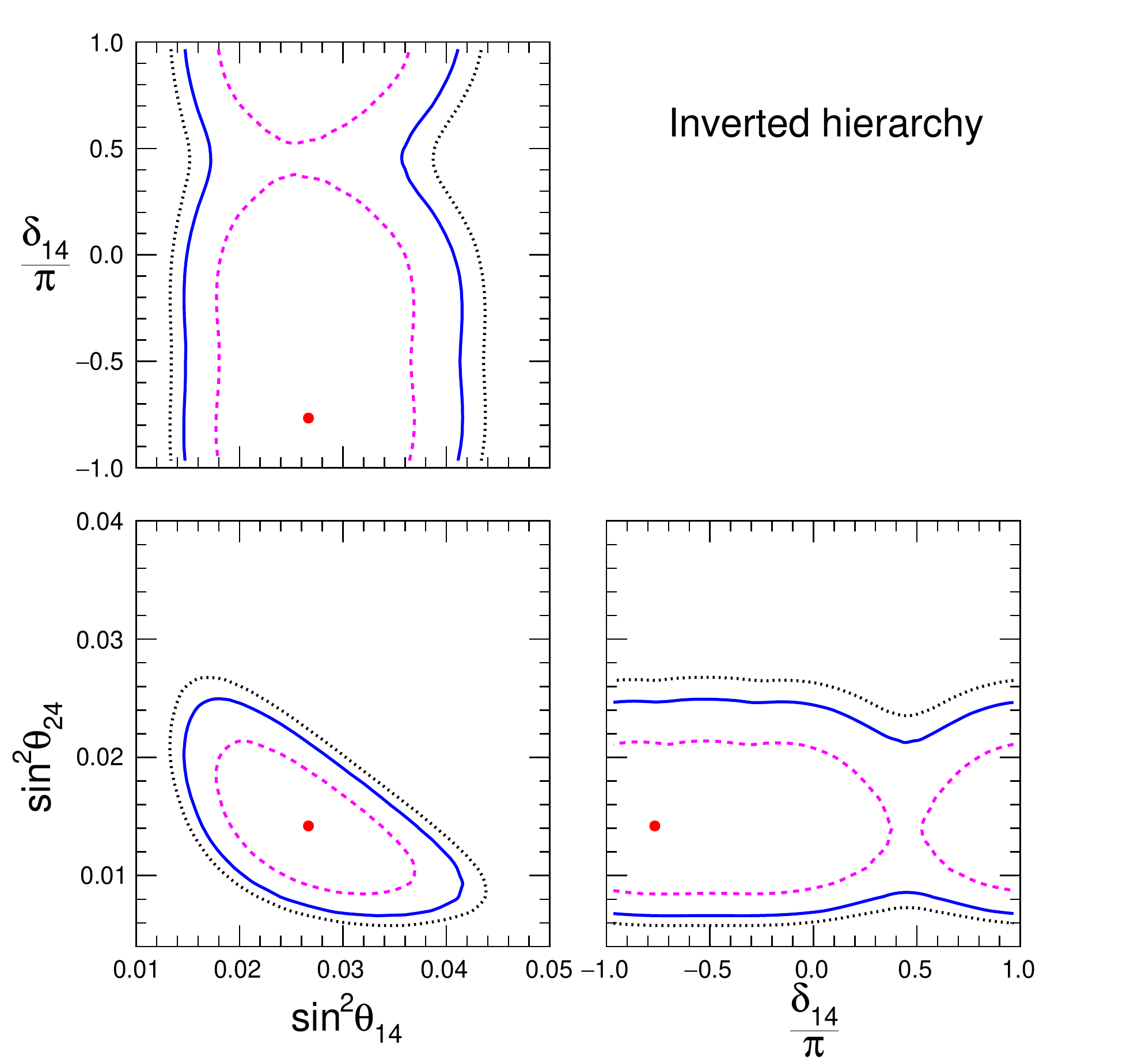}
\vspace*{0.2cm}
\caption{Regions allowed by the combination of the SBL and LBL data (NO$\nu$A and T2K)
together with the $\theta_{13}$-sensitive reactor results. Figure taken from~\cite{Capozzi:2016vac}.
\label{fig:3pan_sbl_lbl_IH}}
\end{figure}  

Figure~\ref{fig:3pan_sbl_lbl_IH}, taken from such a work, displays the 
projections of the $\Delta \chi^2$ for inverted hierarchy (IH). 
The left-bottom panel reports the projection on the plane of the two mixing
angles $(\theta_{14}, \theta_{24})$. The other two panels display the 
constraints in the plane formed by each one of the two mixing angles and the 
new CP-phase $\delta_{14}$. Similar results (not shown) are obtained for normal hierarchy (NH). 
The three contours are drawn for $\Delta\chi^2=2.3,\,\, 4.6,\,\, 6.0$,
corresponding to 68\%, 90\% and 95\% C.L.  for 2 d.o.f.
The overall goodness of fit is acceptable (${\mathrm {GoF}} = 24\%$), while the parameter goodness of fit, 
which measures the statistical compatibility between the appearance and
disappearance data sets, is lower (${\mathrm {GoF}} = 7\%$). This implies that even if the 
closed contours presented for the two new mixing angles $\theta_{14}$ and $\theta_{24}$
exclude the 3-flavor case with high significance (slightly more than six standard deviations), 
one cannot naively interpret this circumstance as an evidence for sterile neutrinos. In addition,
we recall that light sterile neutrinos, unless endowed with new properties, 
are in tension with cosmological data.

\begin{figure}[t!]
\hspace*{0.1cm}
\centerline{
\includegraphics[width=1.2\textwidth]{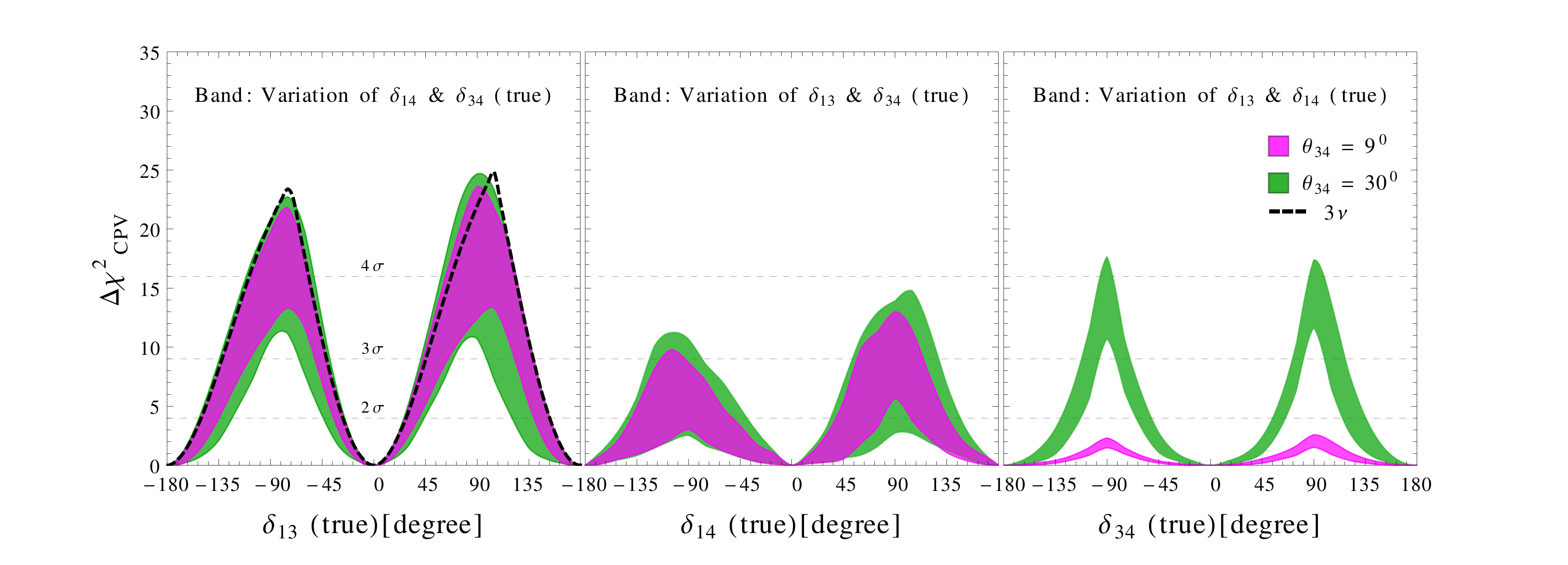}
}
\caption{
DUNE discovery potential of the CPV induced by the three CP phases
involved in the 3+1 scheme. See the text for details.
Figure taken from~\cite{Agarwalla:2016xxa}.}
\label{fig:CPV_4nu}
\end{figure}

Given that NO$\nu$A and T2K posses already a weak sensitivity to the new CP-phase $\delta_{14}$,
it is very interesting to ask how things will improve at the planned LBL experiments.
This issue has been investigated in detail in the works~\cite{Agarwalla:2016mrc,Agarwalla:2016xxa,Agarwalla:2016xlg}
(see also~\cite{Berryman:2015nua,Dutta:2016glq}).
In Fig.~\ref{fig:CPV_4nu}, taken from~\cite{Agarwalla:2016xxa},
we provide an illustrative example concerning the DUNE experiment.
 The bands displayed in the left, middle and right panels represent the discovery potential 
of the CPV induced, respectively, by $\delta_{13}$, $\delta_{14}$ and $\delta_{34}$. 
The thinner (magenta) bands correspond to the case in which all the three 
new mixing angles have the same value $\theta_{14} = \theta_{24} 
= \theta_{34} = 9^0$. The thicker (green) bands correspond to the situation in which 
$\theta_{14} = \theta_{24} = 9^0$ and $\theta_{34} = 30^0$. 
In each panel, the bands were obtained by varying the true values of
the two undisplayed CP-phases in the range $[-\pi,\pi]$.
In all cases, marginalization over the hierarchy was performed with NH as true choice. From
Fig.~\ref{fig:CPV_4nu} we learn that the sensitivity to $\delta_{14}$ will substantially
increase at DUNE at the price of loosing some information on the standard CP phase
$\delta_{13}$.

\section{Conclusions}
 
We have presented a concise discussion of the current phenomenology of light 
sterile neutrinos. 
In the case of a discovery at a new short-baseline experiment we will face the challenge
of determining all the parameters that govern the extended framework and in particular
the new CP-violating phases. In this context LBL experiments can
give an important contribution being sensitive to new CP-violation phenomena. 
Therefore, the two classes of experiments (SBL and LBL) will be synergic in the
searches of sterile neutrinos.

\Acknowledgements
I would like to thank the organizers for the very stimulating and enjoyable 
conference. This work is supported  by the project 
{\em Beyond three neutrino families} within the FutureInResearch program, Fondo di Sviluppo e Coesione 2007-2013, APQ Ricerca Regione Puglia ÒProgramma regionale a sostegno della specializzazione intelligente e della  sostenibilit\`a sociale ed ambientale. 
I also acknowledge support by the research project 
{\em TAsP} funded by the Instituto Nazionale di Fisica Nucleare (INFN).


\begin{thebibliography}{99}



\bibitem{Palazzo:2013me} 
  A.~Palazzo,
  Mod.\ Phys.\ Lett.\ A {\bf 28}, 1330004 (2013).

  
\bibitem{Klop:2014ima}
  N. Klop and A. Palazzo,
  {Phys.  Rev. D} {\bf 91} no. 7, 073017 (2015).


\bibitem{Palazzo:2011rj} 
  A.~Palazzo,
  Phys.\ Rev.\ D {\bf 83}, 113013 (2011).

\bibitem{Palazzo:2012yf} 
  A.~Palazzo,
  Phys.\ Rev.\ D {\bf 85}, 077301 (2012).
  
  
\bibitem{Giunti:2009xz} 
  C.~Giunti and Y.~F.~Li,
  Phys.\ Rev.\ D {\bf 80}, 113007 (2009).


\bibitem{Nunokawa:2003ep} 
  H.~Nunokawa, O.~L.~G.~Peres and R.~Zukanovich Funchal,
  Phys.\ Lett.\ B {\bf 562}, 279 (2003).




\bibitem{Athanassopoulos:1996jb} 
  C. Athanassopoulos {\it et al.}  [LSND Collaboration],
  {Phys. Rev. Lett.}  {\bf 77} 3082 (1996). 
  
 
\bibitem{Armbruster:2002mp} 
  B. Armbruster  {\it et al.}  [KARMEN Collaboration],
  {Phys.\ Rev.\ D} {\bf 65} 112001 (2002).
   
      
\bibitem{AguilarArevalo:2012va} 
  A. A. Aguilar-Arevalo {\it et al.},  
  arXiv:1207.4809 [hep-ex].

   
  
\bibitem{Antonello:2012pq} 
  M. Antonello {\it et al.}
  {Eur. Phys. J. C} {\bf 73} no. 3, 2345 (2013).
    
  
\bibitem{Agafonova:2013xsk} 
  N. Agafonova {\it et al.}  [OPERA Collaboration],
  {JHEP} {\bf 1307},  004 (2013) 
  [Addendum-ibid.\  {\bf 1307} 085].


\bibitem{Palazzo:2015wea} 
  A. Palazzo,
  {Phys. Rev. D} {\bf 91} no. 9, 091301 (2015). 


\bibitem{Mueller:2011nm} 
  T. A. Mueller {\it et al.},
  {Phys. Rev. C} {\bf 83} 054615 (2011).
  
\bibitem{Huber:2011wv} 
  P. Huber,
  {Phys. Rev. C} {\bf 84} 024617 (2011) 
  [2012 Erratum-ibid. {\bf 85} 029901 (2011)].




\bibitem{Mention:2011rk} 
  G. Mention {\it et al.} 
  {Phys.\ Rev.\ D} {\bf 83} 073006 (2011). 


\bibitem{An:2016srz} 
  F.~P.~An {\it et al.} [Daya Bay Collaboration],
  Chin.\ Phys.\ C {\bf 41}, no. 1, 013002 (2017).

\bibitem{Schoppmann:2016iww} 
  S.~Schoppmann [Double Chooz Collaboration],
  PoS HQL {\bf 2016}, 010 (2017).

\bibitem{Seo:2016uom} 
  S.~H.~Seo {\it et al.} [RENO Collaboration],
  arXiv:1610.04326 [hep-ex].

\bibitem{Ko:2016owz} 
  Y.~J.~Ko {\it et al.},
  Phys.\ Rev.\ Lett.\  {\bf 118}, no. 12, 121802 (2017).
 

\bibitem{Palazzo:2013bsa}
  A. Palazzo,
 {JHEP} {\bf 1310}, 172 (2013). 

  

\bibitem{Giunti:2016elf} 
  C.~Giunti,
  Phys.\ Lett.\ B {\bf 764}, 145 (2017).
  
\bibitem{An:2017osx} 
  F.~P.~An {\it et al.} [Daya Bay Collaboration],
  [arXiv:1704.01082 [hep-ex]].

 
\bibitem{Abdurashitov:2005tb}
  J. N. Abdurashitov  {\it et al.},
  {Phys. Rev.  C}, {\bf 73} (2006) 045805.
    
  

  
\bibitem{Gariazzo:2017fdh} 
S.~Gariazzo, C.~Giunti, M.~Laveder and Y.~F.~Li,
arXiv:1703.00860 [hep-ph].
  


\bibitem{Palazzo:2015gja} 
  A. Palazzo,
  {Phys.  Lett.  B}, {\bf 757} 142 (2016).
    



\bibitem{Adamson:2017gxd} 
  P.~Adamson {\it et al.} [NOvA Collaboration],
  arXiv:1703.03328 [hep-ex].


\bibitem{Abe:2017uxa} 
  K.~Abe {\it et al.} [T2K Collaboration],
  Phys.\ Rev.\ Lett.\  {\bf 118}, no. 15, 151801 (2017).

    


\bibitem{Capozzi:2016vac} 
  F.~Capozzi, C.~Giunti, M.~Laveder and A.~Palazzo,
  Phys.\ Rev.\ D {\bf 95}, no. 3, 033006 (2017).


  


\bibitem{Agarwalla:2016mrc} 
  S.~K.~Agarwalla, S.~S.~Chatterjee, A. Dasgupta and A. Palazzo,
  JHEP {\bf 1602}, 111 (2016).

\bibitem{Agarwalla:2016xxa} 
  S.~K.~Agarwalla, S.~S.~Chatterjee and A.~Palazzo,
  JHEP {\bf 1609}, 016 (2016).
  
\bibitem{Agarwalla:2016xlg} 
  S.~K.~Agarwalla, S.~S.~Chatterjee and A.~Palazzo,
  Phys.\ Rev.\ Lett.\  {\bf 118}, no. 3, 031804 (2017).
   
\bibitem{Berryman:2015nua} 
  J.~M.~Berryman, A.~de Gouvea, K.~J.~Kelly and A.~Kobach,
  Phys.\ Rev.\ D {\bf 92}, no. 7, 073012 (2015).
     
\bibitem{Dutta:2016glq} 
  D.~Dutta, R.~Gandhi, B.~Kayser, M.~Masud and S.~Prakash,
  JHEP {\bf 1611}, 122 (2016).
     
\end{thebibliography}
\end{document}